\documentclass[11pt,aps,groupedaddress,preprintnumbers,
               nofootinbib]{revtex4}

\usepackage{graphicx}  

\def\arnps#1#2#3{{\it Ann.\ Rev.\ Nucl.\ Part.\ Sci.} {\bf #1}, #2 (20#3)}
\def\mpla#1#2#3{{\it Mod.\ Phys.\ Lett.} {\bf A#1}, #2 (20#3)}
\def\npb#1#2#3{{\it Nucl.\ Phys.} {\bf B#1}, #2 (20#3)}
\def\plb#1#2#3{{\it Phys.\ Lett.} {\bf B#1}, #2 (20#3)}
\def\oldplb#1#2#3{{\it Phys.\ Lett.} {\bf B#1}, #2 (19#3)}
\def\prd#1#2#3{{\it Phys.\ Rev.} {\bf D#1}, #2 (20#3)}
\def\oldprd#1#2#3{{\it Phys.\ Rev.} {\bf D#1}: #2 (19#3)}
\def\prep#1#2#3{{\it Phys.\ Rept.} {\bf #1}, #2 (20#3)}
\def\prl#1#2#3{{\it Phys.\ Rev.\ Lett.} {\bf #1}, #2 (20#3)}
\def\oldprl#1#2#3{{\it Phys.\ Rev.\ Lett.} {\bf #1}, #2 (19#3)}

\def\sss{\scriptscriptstyle}	

\def\thingie{\hbox{\kern-9pt\raise1pt%
         \hbox{{\fiverm(}{\lower1.5pt\hbox{\twelvebf--}}{\fiverm)}}}}
\def\pmdiff#1#2{\raise.5ex\hbox{$\sss +#1}$%
    \kern-2.8em\lower1ex\hbox{${\sss-#2}$}} 

\def\barp{{\raise.35ex\hbox{${\sss (}$}}---{\raise.35ex\hbox{${\sss )}$}}}				
\def\bdbarp{\hbox{$B_d$\kern-1.4em\raise1.4ex\hbox{\barp}}}
\def\nlpbarp{\hbox{$\nu_{\ell^{\prime}}$\kern-1.4em \raise1.4ex\hbox{\barp}}}

\def\decayarrow{\kern0.2em\hbox{$\raise1.08ex\hbox{\big|}\kern-0.5em
                \longrightarrow$}}

\def\lsim{\;\raisebox{-.6ex}{$\stackrel{<}{\sim}$}\;}

\def\ra{\rightarrow}

\newcommand{\Eq}[1]{Eq.~(\ref{eq#1})}
\newcommand{\beq}{\begin{equation}}
\newcommand{\eeq}{\end{equation}}

\begin{document}

\title{Leptogenesis at the Electroweak Scale}

\author{Boris Kayser}
\email{boris@fnal.gov}
\affiliation{\it Theoretical Physics Department, Fermilab, P.O. Box 500, Batavia, IL 60510  USA}
\author{Gino Segre}
\email{segre@dept.physics.upenn.edu}
\affiliation{\it Department of Physics \& Astronomy, University of Pennsylvania, Philadelphia, PA 19104 USA}
\date{\today}

\preprint{FERMILAB-PUB-10-431-T}

\begin{abstract}
In this note we propose a model of leptogenesis in which the scale for the
mass of the necessary heavy neutral lepton is similar to the scale of electroweak symmetry
breaking.
\end{abstract}

\maketitle


\section{Introduction} \label{Intro}

Leptogenesis \cite{ref1}-\cite{ref3} appears to provide a natural explanation of the cosmic baryon-antibaryon asymmetry. In leptogenesis, CP-violating decays of heavy Majorana neutrinos produce a lepton-antilepton asymmetry, and then sphaleron processes at and above the electroweak symmetry breaking scale convert part of this asymmetry into the observed baryon-antibaryon asymmetry. The heavy neutrinos are see-saw partners of the observed light ones. In the standard type-I see-saw picture, one and the same matrix  of Yukawa coupling constants leads to the CP-violating decays of the heavy neutrinos, to the Dirac masses of the light neutrinos, and to all CP-violating effects among the Standard Model leptons. This linking of diverse physical phenomena is an attractive feature of leptogenesis, and of the see-saw picture from which it springs. However, this linking also leads to an important constraint: if heavy neutrino decay is to provide the degree of CP violation needed to explain the observed baryon-antibaryon asymmetry, and in addition light neutrino masses of the observed order of magnitude are to be obtained, then the heavy neutrinos must have masses of $10^{(8-9)}$ GeV or more \cite{ref4}, putting them far out of reach of current or foreseeable accelerators.

In this paper, we propose a new version of leptogenesis in which the heavy neutrinos have masses of the order of the electroweak scale, (100 -- 200) GeV. This puts them well within reach of the Large Hadron Collider \cite{ref5}. Our proposal is not without its drawbacks, and the heavy neutrinos may prove difficult to observe despite their low masses. However, we believe the scheme is interesting enough to warrant serious consideration, and hope that this paper will stimulate that.

In either ``standard'' leptogenesis or the alternative being proposed here, the heavy neutrinos must decay out of equilibrium. That is, when a heavy neutrino $N$, with mass $m_N$, decays to a Standard Model (SM) lepton $L$ and a Higgs boson $\phi$ via a Yukawa coupling constant $y$, then the $N$ decay rate $\Gamma_D \sim (y^2 / 8\pi) \, m_N$ must not exceed the Hubble expansion rate of the universe, $H$, when the temperature $T$ is $m_N$. Since 
\beq
H(T=m_N) = 1.66  \sqrt{g^*}  \left. \frac{T^2}{m_{Pk} } \right| _{T=m_N} ~~,
\label{eq1}
\eeq
where $g^* \sim 100$ is the number of relativistic degrees of freedom, and $m_{Pk} \sim 10^{19}$ GeV is the Planck mass, we require that
\beq
y^2 \lsim 400 \frac{m_N}{m_{Pk}} ~~.
\label{eq2}
\eeq
In the see-saw picture \cite{ref6}, the masses $m_\nu$ of the light neutrinos are related to the masses $m_N$ of their heavy see-saw partners by a relation of the form
\beq
m_\nu \sim \frac{(vy)^2}{m_N} ~~,
\label{eq3}
\eeq
where $v = 174$ GeV is the vacuum expectation value of the SM neutral Higgs field. Combining Eqs. (\ref{eq2}) and (\ref{eq3}), we see that leptogenesis requires that 
\beq
m_\nu \lsim 10^{-3} \mathrm{eV} ~~.
\label{eq4}
\eeq
Interestingly, the light neutrino masses do come within a few orders of magnitude of satisfying this approximate relation, which is generic to leptogenesis models.

\begin{figure}[hbt]
\begin{center}
\includegraphics[scale=0.20]{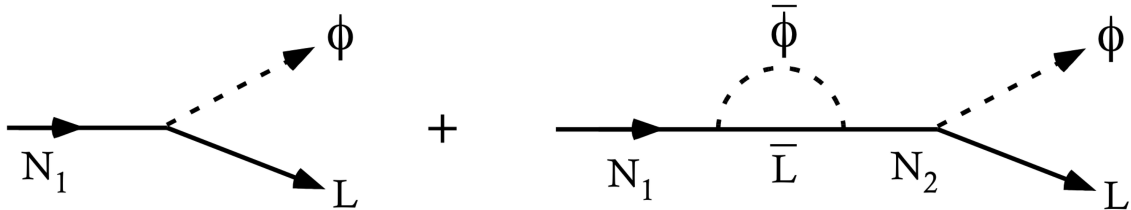}
\caption{The tree diagram and an illustrative loop diagram whose interference can lead to leptogenesis.}
\label{f1}
\end{center}
\end{figure}

In the heavy neutrino decays that drive leptogenesis, the CP violation that is needed to produce a matter-antimatter asymmetric universe arises from interference between a dominating tree-level decay diagram and various loop diagrams. Suppose there are three heavy neutrinos $N_i, \: i=1, 2, 3$. In the standard version of leptogenesis, the tree diagram for the decay $N_1 \ra L\phi$ of the lightest heavy neutrino $N_1$, and one of the loop corrections to this decay, are the diagrams shown in Fig. \ref{f1}. Let us call the 
Yukawa coupling constants at the various vertices in these diagrams generically $y$. For the CP-violating asymmetry produced by the interference between the diagrams,
\beq
\epsilon_{CP} \equiv \frac{\Gamma(N_1 \ra L\phi) - \Gamma(N_1 \ra\bar{ L}\bar{\phi})}{\Gamma(N_1 \ra L\phi) +  \Gamma(N_1 \ra\bar{ L}\bar{\phi})} ~~,
\label{eq5}
\eeq
we have
\beq
\epsilon_{CP} \approx \, \frac{1}{\pi}   \frac{{\cal O} (|y|^4)}{{\cal O} (|y|^2)}   \,  \eta~~,
\label{eq6}
\eeq
where $\eta$ is a factor parametrizing the CP-violating phases. As this illustrates, in standard leptogenesis, the CP-violating asymmetry $\epsilon_{CP}$ arising from heavy neutrino decays is of 2nd order in the Yukawa coupling constants. More specifically, $\epsilon_{CP} \sim |y|^2 / (10 \mathrm{\; to\; } 100)$\cite{ref2}. Since $\epsilon_{CP}$ must be $\sim 10^{-6}$ in order for leptogenesis to yield the observed baryon to photon number density ratio of the universe\cite{ref3}, this implies that $y^2$ must be in the range $10^{-4} - 10^{-5}$. This range seems very reasonable, given prevailing prejudices regarding the acceptable values of coupling constants.

From cosmological observations and tritium $\beta$ decay experiments, we know that the light neutrino masses lie in the eV range or below. From neutrino oscillation experiments, we know that at least one light neutrino has a mass of 0.04 eV or above. Therefore, we assume the light neutrino masses $m_\nu$ to be of order $10^{-1}$ eV. If the see-saw relation, \Eq{3}, is to yield light neutrino masses of this order when $y^2 \sim 10^{-(4-5)}$, the heavy neutrinos must have masses $m_N \sim 10^{(9-10)}$ GeV\cite{ref3}. Thus, in standard leptogenesis, the heavy neutrinos are very far beyond the range of any present or foreseeable particle accelerator. In addition, they raise the question of what physics generates their $10^{(9-10)}$ GeV mass scale.

We would like to present a novel version of leptogenesis in which the heavy neutrinos have masses that are at the electroweak scale. This puts them kinematically within reach of the Large Hadron Collider, and eliminates the need for a new high-mass scale of unknown origin. Our model hinges on Higgs boson quartic couplings.

Assuming that the light neutrino masses are still generated by the see-saw mechanism, we see from the see-saw relation, \Eq{3}, that if $m_N$ is only at the electroweak scale, then the coupling $y^2$ must be quite small. In particular, if $m_N \sim 200$ GeV, then $y^2 \sim 10^{-12}$. The out-of-equilibrium condition, \Eq{2}, requires a somewhat smaller coupling, $y^2 \lsim 10^{-14}$. While such a coupling constant is indeed small, we note that the Yukawa coupling constant $g_{ee}$ that is generally thought to lead to the electron's mass is not markedly larger: $g^2_{ee} \sim 10^{-11}$.

As we have noted, standard leptogenesis requires that $y^2 \sim 10^{-(4-5)}$ in order that the CP-violating asymmetry $\epsilon_{CP}$ produced by $N$ decays be sufficiently large. Thus, in standard leptogenesis, the Yukawa coupling $y^2 \sim 10^{-(12-14)}$ appropriate to our alternative scenario would be far too small. However, as we shall see, in this new scenario, $\epsilon_{CP}$ is actually independent of $y$.

\section{The electroweak-scale scenario} \label{Sect2}

We will assume that there are three  SU(2) X U(1) scalar doublets,
\beq
\phi_a = \left( \begin{array}{c}
	\phi_a^+  \\  \phi_a^0  \end{array}    \right) ~~,~~a=1,2,3~~,
\label{eq7}
\eeq
that couple to quarks and leptons.
We will also assume that the potential is such that $\phi^0_1$ acquires a nonvanishing vacuum expectation value, $<\phi^0_1>\;  \equiv v_1 \neq 0$, but the vacuum expectation values of all the other scalar fields vanish. That is, $<\phi^0_a>\; \equiv v_a = 0$ for $a$ = 2, 3. The potential will naturally lead to these vacuum expection values (vevs) when only the $\phi_1$ mass term is negative, and there are no terms linear in $\phi_2$ or $\phi_3$.

The Yukawa interactions that are of primary interest to us in the consideration of leptogenesis are the ones that couple the scalar fields to leptons. Those interactions are given by
\begin{eqnarray}
-{\cal L}_{{\mathrm{Yukawa}}} &  = &\phantom{+} g_a^{\alpha i} (\overline{\nu_{\alpha L}} \, \phi_a^+ + \overline{\ell_{\alpha L}} \, \phi_a^0) \ell_{iR}  \nonumber  \\
&  & + \,y_a^{\alpha}  (\overline{\nu_{\alpha L}}\, \overline{ \phi_a^0} - \overline{\ell_{\alpha L}} \, \phi_a^-) N_R ~~.
\label{eq8}
\end{eqnarray}
(Summation over repeated indices is assumed.) Here, $\nu_{\alpha L}$ and $\ell_{\alpha L},\, \alpha=e,\mu,\tau$, are, respectively, the neutrino and charged lepton of the Standard Model left-handed lepton doublets. Similarly, $ \ell_{iR}$ and $N$, are, respectively, the charged and neutral right-handed electroweak-singlet leptons. We note that in the conventional model of leptogenesis, as illustrated in Figure \ref{f1}, at least two  massive singlet $N$ fields are necessary to obtain a non-vanishing effect while, as we shall show, one is sufficient in our case.

\begin{figure}[hbt]
\begin{center}
\includegraphics[scale=0.70]{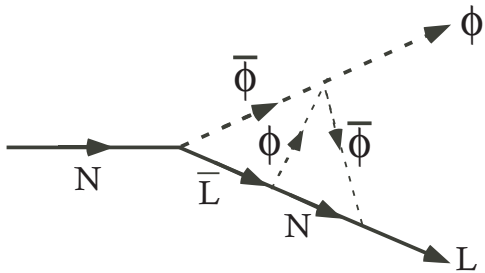}
\caption{Two loop diagram contributing to leptogenesis.}
\label{f2}
\end{center}
\end{figure}

The generic class of diagrams on which we wish to focus is illustrated in Figure~\ref{f2}.
These are not taken into account in the conventional estimates of leptogenesis because their contributions are smaller than those of Figure~\ref{f1} by a factor of $\lambda/ 4\pi^2$, where $\lambda$ is a generic four scalar field coupling constant and such constants are usually taken to be ${\cal O}(10^{-2})$ in order for perturbation theory to be meaningful in calculating Higgs potentials.

In our model, we assume that the scalar mesons and the $N$ all have comparable masses that are of the order of the electroweak symmetry breaking scale. We assume that the $N$ acquires its mass through a Majorana mass term which serves as our source of lepton-number nonconservation. The scalar doublet $\phi_1$ will of course not acquire mass until after the symmetry breaking has occurred, while $\phi_2$ and $\phi_3$, with positive mass terms in the Lagrangian, will already be massive before such symmetry breaking has occurred.

\begin{figure}[hbt]
\begin{center}
\includegraphics[scale=0.70]{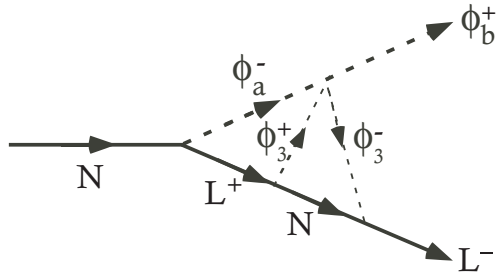}
\caption{Significant two loop diagram contributing to leptogenesis in this model.}
\label{f3}
\end{center}
\end{figure}

We will choose our masses to be ordered such that the third scalar doublet has the largest mass, noting that this does not require any fine tuning. The order we select is
\beq
M_3 > M_N > M_{1,2} ~~.
\label{eq9}
\eeq
This means that the decays $N \ra L + \phi_{1,2}$ are allowed, while $N \ra L + \phi_3$ is forbidden. This leads to an interesting possibility. Since $M_3 > M_N$, we need not restrict $y_3$ to be as small as $y_{1,2}$, for $\phi_3$ neither contributes to neutrino masses nor to $N$ decay modes. In fact, there is no reason that $y_3$ cannot be ${\cal O}(1)$. In this case, the diagrams illustrated in Figure~\ref{f3}, a subclass of those shown in Figure \ref{f2}, can give a large contribution to leptogenesis. 

We also note that loop diagrams for which the initial coupling is $N \ra L + \phi_3$ are also of course present but they will not contribute to leptogenesis since $M_3 > M_N$ implies that they have no discontinuity, and such a discontinuity is necessary in order to make a non-zero contribution to leptogenesis.

The quartic couplings of the scalar fields are generally of the form 
\begin{eqnarray}
V  &=&  \sum_{a,b =1,2}( \lambda_{33ab} \,\phi_3^\dagger\ \cdot \phi_3 \,\phi_a^\dagger \cdot \phi_b +	 \lambda_{3a3b} \phi_3^\dagger\phi_a \phi_3^\dagger\phi_b
			+ \lambda_{3ab3} \phi_3^\dagger \cdot \phi_a \,\phi_b^\dagger \cdot \phi_3)		 \nonumber \\
			&& +\; \lambda_{3333} \,\phi_3^\dagger \cdot \phi_3\, \phi_3^\dagger \cdot \phi_3 ~ + \mathrm{h.c.} ~~,
\label{eq10}
\end{eqnarray}
where we have written only those terms in the potential that involve the  $\phi_3$  field. We anticipate that all the $\lambda$  coupling constants are of the same  order of magnitude, with one exception. The quartic couplings involving two $\phi_3$ and two $\phi_1$ fields must be chosen so that $y_3^2 \lambda_{3131} \leq 10^{-12}$. Otherwise neutrinos will acquire unacceptably large masses through the type of one-loop diagrams studied by Ma \cite{ref8} . They are not, however, all real even though the potential is Hermitian.

Generically writing $y$ for $y_{1,2}$, assumed to be comparable in magnitude, we see that diagrams such as the ones of Figure~\ref{f3}, interfering with the tree diagram, make a contribution to the lepton asymmetry of \Eq{5} that is of order
\beq
\epsilon_{CP} \approx \frac{|y|^2 \, |y_3|^2 \, |\lambda _{3a3b}|}{4\pi^3 \, |y|^2} \; \eta^\prime ~~,
\label{eq11}
\eeq
where $\eta^\prime$ is a factor that depends on the detailed evaluation of the diagram, and on CP-violating phases, as in \Eq{6},  but is generally ${\cal O}(1)$. If $M_N$  were appreciably smaller than $M_3$ , $\eta^\prime$ would contain a suppression of order $(M_N / M_3)^2$  that came from the evaluation of the loop diagram in Figure~\ref{f3}, but we have assumed $M_N$ and $M_3$ are comparable in magnitude. The additional $1/4 \pi^2$ in \Eq{11} relative to the analogous \Eq{6} follows because the appropriate diagrams generating $\epsilon_{CP}$ involve two loops rather than one. Since $y_3$  is ${\cal O}(1)$ and $\lambda$ is ${\cal O}(10^{-2})$, we see that the lepton asymmetry can readily reach the desired value of $10^{-6}$.

To illustrate how the CP violation and nonvanishing lepton number actually arise in our model, let us assume that $\phi_3$ couples to $\phi_2$ but not $\phi_1$. Let us also assume that the $N$ mass is such that leptogenesis takes pace at a temperature below 1 TeV, but well above the electroweak phase transition. Then, at the time of leptogenesis, the Standard Model leptons $\ell_\alpha$ and $\nu_\alpha$ will all be massless. With $\phi_3$ coupling to $\phi_2$ but not $\phi_1$, the two-loop diagrams of the kind illustrated in Figure~\ref{f3} will contribute to $N$ decays yielding a $\phi_2$, but not to those yielding a $\phi_1$. Omitting irrelevant factors, the amplitude for the decay $N \ra \ell^-_\alpha + \phi^+_2$, Amp$(N \ra \ell^-_\alpha \phi^+_2)$, is given by
\beq
\mathrm{Amp}(N \ra \ell^-_\alpha \phi^+_2) = y^\alpha_2 + \sum_{\beta = e,\mu,\tau} {y_2^\beta}^* y_3^\beta y_3^\alpha  \lambda^*_{3232} \, K ~,
\label{11.1}
\eeq
where $K$ is a kinematical factor. The first term in this amplitude is from the tree diagram, and the second is from the loop. This amplitude takes into account all the lepton flavors $\beta$, and both the $\ell^+_\beta \phi^-_2$ and $\overline{\nu_\beta} \overline{\phi^0_2}$ configurations, in the intermediate state of the two-loop diagram. Omitting an overall phase space factor, we find from this amplitude that for the charged-lepton final state, the lepton-antilepton difference, including all final lepton flavors $\alpha$, is given by
\beq
\sum_\alpha \Gamma(N \ra \ell^-_\alpha \phi^+_2) - \sum_\alpha \Gamma(N \ra \ell^+_\alpha \phi^-_2) = 4 \Im \left [ (\sum_\alpha {y_2^\alpha} {y_3^\alpha}^* )^2  \lambda_{3232} \right] \, \Im K ~.
\label{11.2}
\eeq
For the neutrino final state, we find that
\beq
\sum_\alpha \Gamma(N \ra \nu_\alpha \phi^0_2) - \sum_\alpha \Gamma(N \ra  \overline{\nu_\alpha} \overline{\phi^0_2} ) = 4 \Im \left [ (\sum_\alpha {y_2^\alpha} {y_3^\alpha}^* )^2  \lambda_{3232} \right] \, \Im K ~.
\label{11.3}
\eeq
That is, we find exactly the same lepton-antilepton difference as in the charged-lepton case, and the two lepton-antilepton differences add. Since the intermediate $\ell^+_\beta \phi^-_2$ or $\overline{\nu_\beta} \overline{\phi^0_2}$ state in the loop diagram can be on shell, $K$  will have a nonvanishing imaginary part, and there is no reason to expect the lepton-antilepton difference to vanish.

To be sure, it is the baryon asymmetry in which we are ultimately interested, and therefore there is the added complication in this picture of having, unlike in the conventional model, the lepton asymmetry and the sphaleron processes that convert this asymmetry into one of baryons occurring at the same scale. Should the generation of the lepton asymmetry at the electroweak scale be occurring too late for the sphaleron processes to effectively convert this asymmetry into one of baryons, one could remedy the situation while maintaining the main features of the model by shifting the $N$ and the third scalar doublet's masses upwards so that the creation of the lepton asymmetry occurred somewhat earlier, say at a scale of 500 GeV. 
We are optimistic that this might not be necessary (see the discussion in Section 3 of \cite{ref2}). It is comforting to note that the magnitude of the lepton asymmetry that is generated in this model is potentially large enough that even some diminution of the conversion is not likely to make its contribution insignificant. 

\section{Experimental Tests}\label{sect3}

An additional attractive feature of such a low scale model of leptogenesis lies in its being in principle testable, unlike the more conventional model in which the neutrino singlet mass is beyond the reach of anticipated accelerators. Experimental tests are foreseeable because one can anticipate that the $\phi_3$ field's coupling to quarks and charged leptons could be large, just like its coupling to $N+\nu$.  In that case, $\phi_3$ could be produced relatively abundantly at a particle accelerator once the energy threshold has been passed.
To be sure, the couplings of  $\phi_3$ to fermions are constrained by the upper limits on neutrino masses. The see-saw expression for these masses, \Eq{3}, may be pictured as arising from a diagram in which $\nu \ra N$ via an interaction with the  $\phi^0_1$ vev, and then $N \ra \nu^c$ via a second interaction with this vev. 
As already noted, if $m_N \sim$ 200 GeV and light neutrino masses of the observed order of magnitude are to be obtained from this see-saw diagram, we must have $y^2_1 \sim 10^{-12}$. Now, neutrino masses can also be induced by a diagram in which the $\nu \ra N$ transition, or the $N \ra \nu^c$ transition, or both, result from the absorption of a $\phi^0_3$ (which does not have a vev) that has come, via a fermion loop, from a $\phi^0_1$ (which does have a vev). 
The fermion-antifermion pair in the loop may be an up-type quark and antiquark, a down-type quark and antiquark, charged leptons, or neutrinos. Considering all possible diagrams of this kind, we find that, given that $y_1$ must be $\sim 10^{-6}, \; y_3$ can be ${\cal O}(1)$ as desired for successful leptogenesis, and the upper bounds on neutrino masses will not be violated, so long as $f(3,j) \, f(1,j) \lsim 10^{-6}$. Here, $j$ runs over all possible fermion-antifermion pairs that can be in the loop, $f(3,j)$ is the coupling of $\phi_3$ to the pair $j$, and $f(1,j)$ is the coupling of $\phi_1$ to this pair. For example, if the loop contains an $e^-e^+$ pair, we have the constraint $g^{ee}_3 g^{ee}_1 \lsim 10^{-6}$ (in the notation of \Eq{8}). This constraint is not at all severe. If $\phi_1$ is the Standard Model Higgs doublet, then $g^{ee}_1 = m_e / v_1 \simeq 3 \times 10^{-6}$. Thus, the coupling of $\phi_3$ to an electron, $g^{ee}_3$, can be ${\cal O}(10^{-1} - 1)$.

At a future electron-positron collider, one might look for e.g. 
\begin{eqnarray}
e^+ + e^- \ra \phi_3^0 \ra \nu +  & \hspace{-1.2cm} N &   \nonumber  \\ 
&   \decayarrow \; \mu^+ +  &  \phi_{1,2}^-      \nonumber  \\ 
&  &   \decayarrow \; \tau^- + \overline{\nu_\tau}
\label{eq12}
\end{eqnarray}
Taking $g^{ee}_3 = 1/3$ and the $\phi^0_3$ mass to be $\sim$ 300 GeV, and estimating the total $\phi^0_3$ width from its principal decay modes, we find that at the peak of the $\phi^0_3$ resonance, $\sigma (e^+ e^- \ra \phi^0_3 \ra \nu N) \sim 2$ nb. This would be a dramatically large cross section.

A similar picture could emerge in a hadron collider, where a comparable process could occur. If, for instance, a down quark and an up anti-quark were to produce a $\phi^-_3$, a possible result might be
\begin{eqnarray}
\bar{u} + d \ra \phi_3^- \ra e^- +  & \hspace{-1.2cm} N &   \nonumber  \\ 
&  \decayarrow \; \mu^- +  & \phi_{1,2}^+     \nonumber  \\ 
&  &  \decayarrow  \;\tau^+ + \nu_\tau
\label{eq13}
\end{eqnarray}
A process such as the one indicated in \Eq{13} would be quite striking. The presence of three charged leptons of different flavors and a neutrino would indicate a new type of physics, since the only reasonable alternative explanation would be leptonic flavor changing neutral currents. The $e + \mu + \tau + \nu$ final state could not come from a pair of Higgs particles with couplings that, as usual, are diagonal in the fermion mass eigenstate basis.

\section{Conclusions}\label{sect4}

We have presented a model for the creation of the cosmic baryon asymmetry via leptogenesis that has some attractive features. The most notable of these is not requiring a significant new mass scale between the one of electroweak symmetry breaking and that of grand unification. We do not claim, of course, that this is the first attempt to achieve such a result. Our model has, as all others, features that may seem contrived, but we believe it is both sufficiently interesting and novel to warrant consideration and perhaps to focus attention once again on leptogenesis at a much lower scale. The model has the not-inconsiderable merit of suggesting experimental tests at colliders.

\section*{Acknowledgments}
It is a pleasure to thank Paul Langacker and Silvia Pascoli for very helpful discussions, Bjorn Garbrecht and Pedro Schwaller for a useful critique of an earlier version of the paper, and a referee for pointing out \cite{ref8} to us. The work of GS was supported by the US Department of Energy under Grant DOE 3071T. Fermilab is operated by Fermi Research Alliance, LLC under Contract No. DE-AC02-07CH11359 with the Department of Energy.

\end{document}